Short Paper

# Seminar and Training Programs Recommender System for Faculty Members of Higher Education Institution

Albert V. Paytaren
College of Engineering and Computing Sciences
Batangas State University

**Abstract**

*Purpose* – This study aims to develop a personalized Recommender System that helps to address the problems encountered by the faculty members of Higher Education Institutions in the selection of Seminar and Training Programs (STP).

*Method* – The researcher used the Descriptive Developmental Method of research to gather information relevant to the current problems and challenges encountered and used these to develop software that addresses the identified challenges. For the development of the software, the researcher adopted a step-wise approach defined in the Incremental Developmental Model. The level of acceptance of the developed system was evaluated by 24 faculty respondents.

*Results* – The level of acceptance of the developed system was classified into functionality, reliability, and usability and the study garnered an evaluation score of 4.65, 4.67, and 4.67 respectively. The overall interpretation of the results of the evaluation is Highly Acceptable.

*Conclusion* – The study created a system that provides seminars and training program recommendations. The developed recommender system was rated Highly Acceptable, respondents were very satisfied with the features of the system and agreed that it was functional, reliable, and usable.

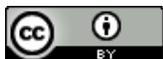

## INTRODUCTION

Many academic institutions considered faculty development as a priority as it helps to improve the academic programs. This also mobilizes the needs of emerging faculty, program, students, and other stakeholders. Most of that teaching personnel in Higher Education Institution (HEI) have come from their previous careers as industry practitioners and as teachers and managers of learning, they have little, or if any, a least formal professional training or experience about the management of learning. The Commission on Higher Education (CHED) Faculty Development Program (FDP) seeks to upgrade the academic qualifications of HEI faculty to enhance faculty performance through continuing professional education, seminar and training programs, workshops, conferences, symposia, and alike. Today, academic practitioners are required to engage in continuous professional, technical, and personal growth due to constant change in their roles and responsibilities. Opportunities for faculty to be improved are essential to organize their teaching and learning materials effectively.

Nowadays, the availability of faculty development programs such as Seminar and Training Programs were abundant both online and offline publishes. But Faculty members consume much time in looking at applicable FDP for them. The challenge of the faculty members is to ensure that they are continuously improving in terms of professional and personal development. Therefore, it is very useful to develop a recommender system that will provide recommendations for Seminar and Training Programs for every faculty member of the HEI.

The objective of this paper is to develop a Seminar and Training Program Recommender System, which has practical significance for every faculty member. It will be designed to address the identified problems encountered by the faculty members in terms of selection of Seminar and Training Programs (STP). To assess its impact, the level of acceptance in terms of functionality, reliability, and usability of the developed software will also be evaluated by selected faculty members of different academic ranks from the Batangas State University ARASOF Nasugbu.

## LITERATURE REVIEW

### Faculty Development Program

The professional development and academic effectiveness of teaching faculty members are connected to its educational vivacity. This can be materialized by a dynamic and active FDP that has been shown to lead the enhancement of faculty's skills in all the five desired domains, i.e., teaching, assessment, curriculum support, organizational leadership, and mentoring (Guraya & Chen, 2017). Faculty development endorses the educational improvements and strategies that are dignitary and are executed professionally. Professional organizations and experts have recommended FDP's for greater awareness and attainment of knowledge in teaching and learning (Ghazvini, Mohammadi, & Jalili, 2014).

### Recommender Systems

A recommendation system uses the large volume of data in the form of text and sentiments available for summarization purposes to make serious and valid decisions (Sardar et al., 2017). Recommender systems are a relatively new area of research in machine learning. There are two main ways that recommender systems produce a list of recommendations for a user, the collaborative and content-based filtering. Collaborative filtering uses past behavior (items that a user previously viewed or purchased, in addition to any ratings the user gave those items) and similar decisions made by other users to create a model. This model then predicts items that the user may find interesting. In content-based filtering, the model uses a series of discrete characteristics of an item to recommend additional items with similar properties (Mohamed, Khafagy, & Ibrahim, 2019). Recommender systems are usually classified based on how recommendations are made. Collaborative filtering makes recommendations based on items owned by users whose taste is similar to those of the given user (Rodriguez & Ferreira, 2016).

The researcher carefully chose the above-mentioned literatures to identify the significance of Faculty Development Programs in Higher Education Institutions. The reviews about the nature of recommender systems also help the researcher to improve the output of this study.

## METHODOLOGY

The researcher used the Descriptive Developmental Method of research to gather information relevant to the current problems and challenges encountered and used these to develop software that addresses the identified challenges. In the conduct of data gathering, the researcher used different forms, manuals and reports such as ISO Forms for Request for Attendance to Seminar, Training Development Plan, Personal Data Sheet (PDS) and Faculty Inventory Forms that are commonly used by the State Universities and Colleges (SUC's) for comprehensive analyzation of the faculty development programs and interests. The researcher comes up with the development of seminar and training programs recommender system.

**Software Development**

For the software development, the researcher utilized the Incremental Model whereas this model combines elements of the linear sequential model (applied repetitively) with the iterative philosophy of prototyping. The incremental process model, like prototyping and other evolutionary approaches, is iterative (Kute & Thorat, 2014). In the incremental model, the requirements are broken down into multiple standalone modules of the software development life cycle. This was done through steps from analysis design, implementation, testing/verification, maintenance (Guru99, 2020). Instead of dividing the Software Development Life Cycle (SDLC) into static, isolated steps, the whole process can instead be designed, tested, and implemented one fraction at a time, successive stages. This feedback will provide valuable input in the total increment of the process and so forth. With each ongoing increment, the prototype is continuously tested, according to identified output and expectations from the user (Thakur, Singh, & Chaudhary, 2015). Developing systems through incremental release requires first providing essential operating functions, then providing system users with improved and more capable versions of a system at regular intervals (Scacchi, 2001).

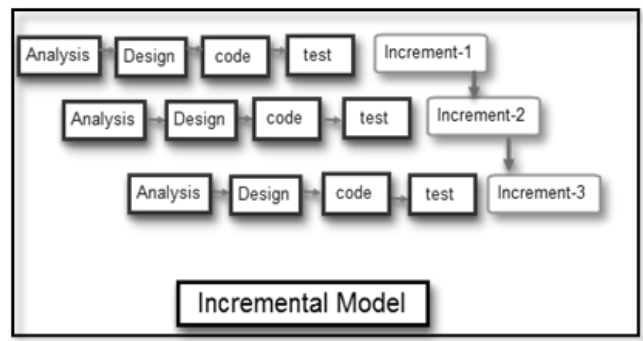

*Figure 1.* Incremental Life Cycle Model

Figure 1 shows the Incremental Life Cycle wherein the system is put into production when the first increment is delivered. The first increment is often a core product where the basic requirements are addressed (Guru99, 2020) such as the account was able to be personalized by each faculty user, and supplementary features were added in the next increments such as the system was able to provide recommendations for the faculty user.

**System Architecture**

Figure 2 shows the architecture of the system, describing Faculty as the end-user of the developed software. The faculty member has to update and personalize their faculty profile, such as the department or college he was deployed, a field of expertise and field of interests, postgraduate education, and the like. Every login, the recommender system provides the recommendations of available seminar and training programs from trusted websites, which is aligned to the faculty's profile. When the faculty member hit like button, that STP will be added to his liked STP and then it will be fed to other faculty member having the same field of expertise and interests. For the faculty member, the more they liked certain STP, the more similar STP will be recommended to other faculty.

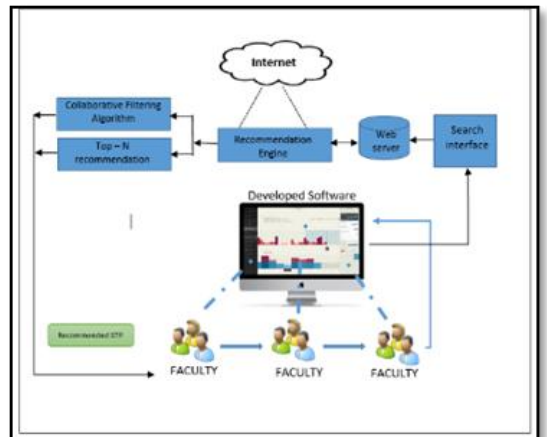

*Figure 2.* The architecture of the System

**Exploded System Architecture**

Figure 3 shows how a collaborative filtering algorithm works in the recommender system. The system initially collects data from the user (faculty member) such as what program he used to teach, the college he was assigned, and the field of expertise and interests he is into. Those data will proceed to clean and the selection and later on will be combined with datasets from a similar faculty member. Based on the data, different seminar and training programs will be analyzed if it matched with the user's preferences.

The researcher used the official website of the Commission on Higher Education (CHED) to look for posted approved Seminar and Training Programs, Conferences, and Training-Workshops.

**Collaborative-Filtering Algorithm**

The prototype was designed to made recommendations based on the profile of the faculty users. Each first-time faculty user should input details relevant to his field of expertise, the college where he was assigned, programs he taught along with his interests. Then, the system will look for faculty with similar data sets of the first-time faculty user, that user data is considered the training dataset since it already contains results – their recommendations generated for the similar faculty.

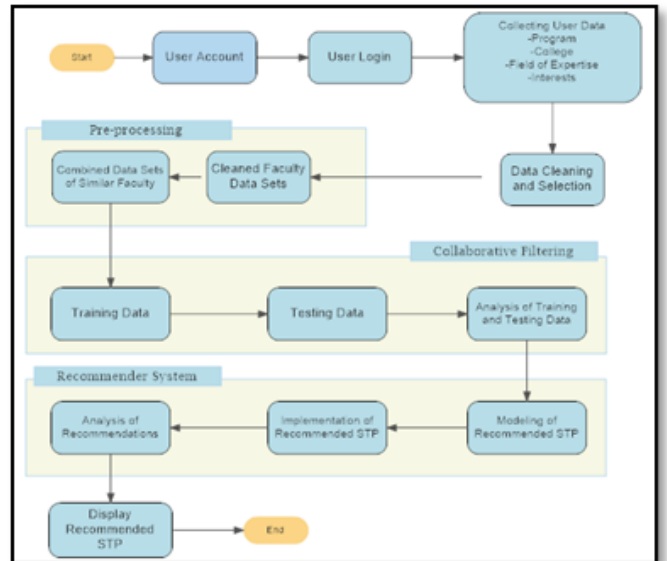

*Figure 3.* The exploded architecture of the System

Then the user data of the first-time faculty user will combine to the user data of other faculty who already use the recommender system. This is now the testing data set, and the algorithm will predict the outcome, based on the historical data for similar faculty. This approach was relatively known as the Collaborative Filtering Algorithm.

Table 1. User Profile of First-time Faculty User

| Name of the Faculty | Josh Magtibay |
|---|---|
| College | College of Accountancy, Business, Economics and International Hospitality Management (CABEIHM) |
| Program | BS-HRM, BS-Accountancy |
| Interests | Accounting, Finance |

Table 2. User Profile of Similar Faculty User

| Name of the Faculty | Benjie A Bautista |
|---|---|
| College | College of Accountancy, Business, Economics and International Hospitality Management (CABEIHM) |
| Program | BS-Accountancy, BS Business Administration |
| Interests | Finance, Entrepreneurship, Business Management |

Table 1 and Table 2 show a different user profile. One is for the first-time faculty user and the other one is the user profile of the similar faculty. Both of them have the same College, but with common details in terms of program and interests.

**Data Gathering Procedure**

Random Sampling was used and 24 out of 240 faculty members were selected as respondents in the study. The researcher used the ranking to determine which among the issues and problems in terms of selecting seminar and training programs were always encountered by the respondents. It involved placing the value in numerical order and assigning new values to indicate where in the ordered set they fall. The respondents gave the smallest number 1 for the most issues and challenges encountered; the next value is 2, to be followed by 3 and so on.

To evaluate the level of acceptance of the developed system, the researcher developed survey questionnaires that include criteria such as Functionality, Reliability, and Usability. Each criterion contains components that characterized the acceptability of the developed system. The respondents put check to the number that corresponds to the user's experience. Weighted Mean was used to determine the functionality, reliability, and usability of the developed system.

## RESULTS

FDP is a significant initiative of HEI that will help to improve and develop every faculty member of the academic institutions. Through the program, there will be continuing professional development intended for the teaching forces. Knowledge, skills, and attitude of the participants will also be a boost. With the use of a personalized STP recommender system, the faculty will avoid problems in the selection of seminars and training programs.

Table 3. Problems challenges encountered by the Faculty in Selection of Seminar and Training Programs

| Issues and Challenges | Mean | Occurrences |
|---|---|---|
| No system recommends available training and seminar programs | 4.84 | Always Encountered |
| Finding Seminar and Training Programs to participate in consumes too much time of the faculty | 4.79 | Always Encountered |
| Available Seminar and Training Program do not apply to the field of specialization and/or needs of the faculty | 4.53 | Always Encountered |
| Seminar and Training Programs posted on the internet are not updated | 4.47 | Always Encountered |
| Seminar and Training Programs posted online and offline sometimes lacks in full details | 4.26 | Always Encountered |

*Legend: 5 - Always Encountered; 4 - Often Encountered; 3 – Sometimes Encountered; 2 – Rarely Encountered; 1 – Never Encountered*

The study was able to rank the identified problems encountered by the faculty in the selection of a Seminar and Training Programs as shown in Table 3. The top issues and challenges encountered the faculty members in terms of selection of available seminar and training programs are "there is no automated system that recommends available training and seminar programs" with a weighted mean of (4.84). All the issues and challenges were given a verbal interpretation of Always Encountered. Tables 4, 5, and 6 show the level of acceptance of the respondents in the developed software in terms of Functionality, Reliability, and Usability.

Table 4. Level of Acceptance of the Respondents in the Developed System in terms of Functionality

| Components of Functionality | Mean | Level of Acceptance |
|---|---|---|
| The system can provide recommendations for updated seminar and training programs | 4.84 | Highly Acceptable |
| The system improves productivity and effectiveness by spending less time in tedious works of finding applicable seminars and training programs for the faculty. | 4.79 | Highly Acceptable |
| The system is capable of recommending seminar and training program based on the faculty's field of expertise | 4.53 | Highly Acceptable |
| The system generates consolidated Seminar and Training Programs attended by the faculty on time | 4.47 | Highly Acceptable |
| The details relative to attended seminar and training programs of a faculty are accessible to the intended user at a given point of time. | 4.26 | Highly Acceptable |
| Composite Mean | 4.65 | Highly Acceptable |

*Legend: 4.21- 5.00 (Highly Acceptable); 3.41- 4.20 (Moderately Acceptable); 2.61- 3.40 (Acceptable); 1.80- 2.60 (Slightly Acceptable) and 1.00- 1.80 (Not Acceptable)*

Table 5. Level of Acceptance of the Respondents in the Developed System in terms of Reliability

| Components of Reliability | Mean | Level of Acceptance |
|---|---|---|
| The system provides an accurate Faculty Profile wherein each faculty has its own account to update their profile. | 4.71 | Highly Acceptable |
| The system is capable of handling errors throughout the process of updating the faculty profile and STP recommendation process. | 4.58 | Highly Acceptable |
| The generated consolidated report in Seminar and Training Programs attended by the faculty is accurate since the system uses automated extraction of data from this assures less human errors | 4.71 | Highly Acceptable |
| Composite Mean | 4.67 | Highly Acceptable |

*Legend: 4.21- 5.00 (Highly Acceptable); 3.41- 4.20 (Moderately Acceptable); 2.61- 3.40 (Acceptable); 1.80- 2.60 (Slightly Acceptable) and 1.00- 1.80 (Not Acceptable)*

Table 6. Level of Acceptance of the Respondents in the Developed System in terms of Usability

| Components of Usability | Mean | Level of Acceptance |
|---|---|---|
| The system offers a convenient user interface that is easy to learn without deliberate effort. | 4.58 | Highly Acceptable |
| The system presents data and information accurately and easy-to-understand manner | 4.79 | Highly Acceptable |
| The generated consolidated seminar and training programs attended by the faculty can serve as input to the top management to a more rationalized utilization of university resources for attendance to training, seminars and similar activities | 4.92 | Highly Acceptable |
| **Composite Mean** | 4.67 | Highly Acceptable |

*Legend: 4.21- 5.00 (Highly Acceptable); 3.41- 4.20 (Moderately Acceptable); 2.61- 3.40 (Acceptable); 1.80- 2.60 (Slightly Acceptable) and 1.00- 1.80 (Not Acceptable)*

**Using the Developed Software**

Figure 4 shows a part of the system that displays the interface of the faculty user were the user profile such as the field of expertise, college, program, and interests of the faculty user were inputted.

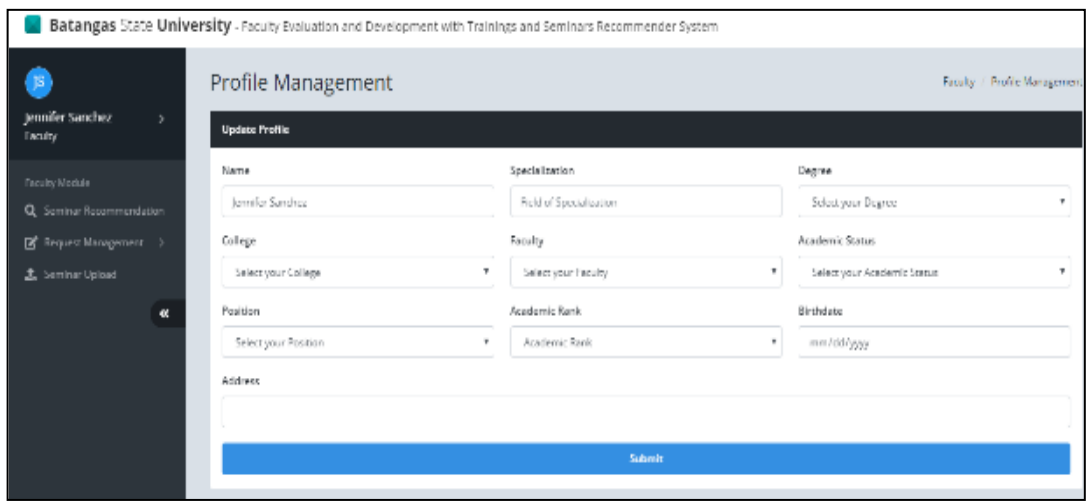

*Figure 4.* Screenshots how user input personal details

**Results of recommendation from the Recommender System**

Figure 5 shows the interface of the faculty user, recommended seminar and training programs were displayed. Faculty user Josh Magtibay of CABEIHM, as shown in Figure 4, has recommended seminars and training programs relevant to his college, program, and interests. It was clear that the displayed recommended seminar and training programs are related to his profile.

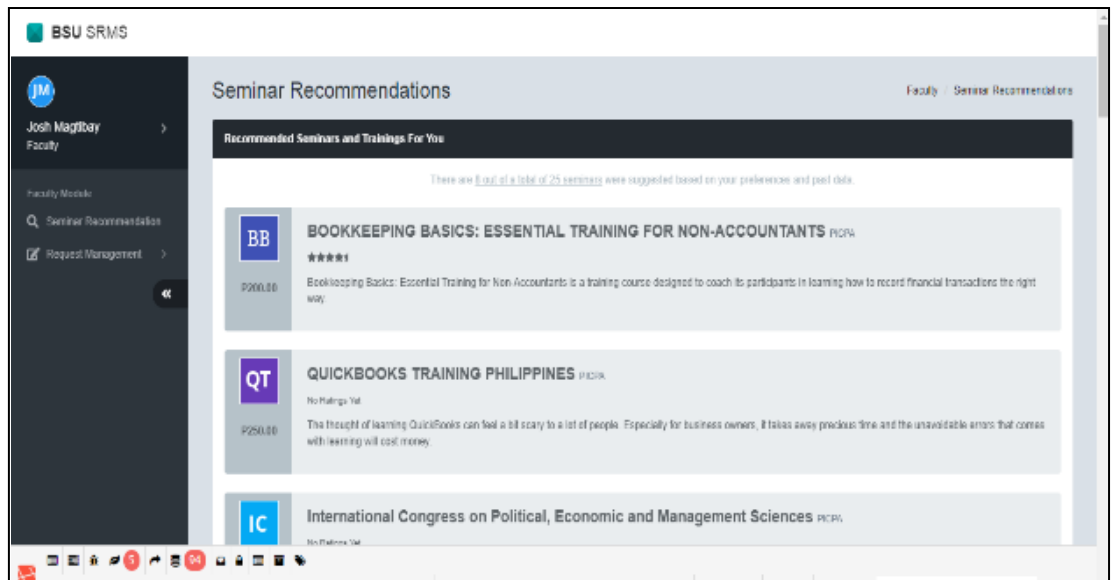

*Figure 5.* Screenshots how System displays recommended STP

## CONCLUSIONS

This paper was able to outline the issues and challenges encountered by the faculty member in the selection of STP. The researcher concluded that doing the selection of STP was done manually and usually consumes too much time and effort. The study created a system that provides seminars and training program recommendations. Since the developed recommender system was rated Highly Acceptable, respondents were very satisfied with the features of the system and agreed that it was functional, reliable, and usable.

## RECOMMENDATIONS

The paper emphatically endorses the use and implementation of the developed recommender system in Higher Education Institutions as it addresses the issues and challenges encountered by the faculty member in the selection of STP.

In this research, the researcher focused on the recommendation of STP for the faculty member of HEI. A key issue that emerges from this study, asking for future research, refers to the effectiveness of the recommendations generated by such a system and how this can be improved using the faculty's reactions to the recommended STP as feedback.